# LightCPPgen: An Explainable Machine Learning Pipeline for Rational Design of Cell Penetrating Peptides

Gabriele Maroni[1], Filip Stojceski[1], Lorenzo Pallante[2], Marco A. Deriu[2], Dario Piga[1], Gianvito Grasso[1]*

*Corresponding author: gianvito.grasso@idsia.ch

[1] Dalle Molle Institute for Artificial Intelligence IDSIA - USI/SUPSI, Via la Santa 1, CH-6962 Lugano-Viganello, Switzerland

[2] PolitoBIOMedLab, Department of Mechanical and Aerospace Engineering, Politecnico di Torino, Corso Duca degli Abruzzi 24, 10129, Torino, Italy.

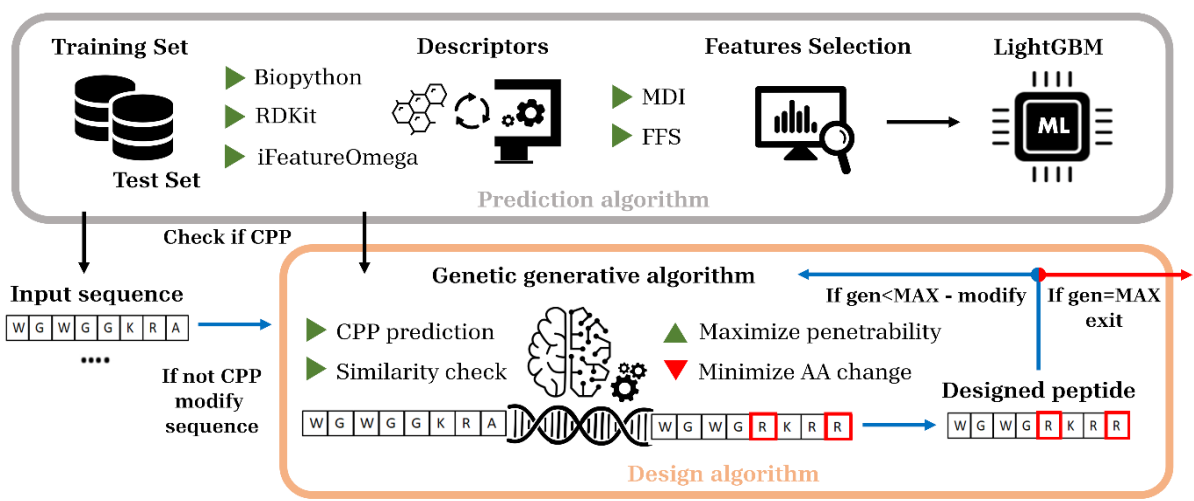

## Abstract

Cell-penetrating peptides (CPPs) are powerful vectors for the intracellular delivery of a diverse array of therapeutic molecules. Despite their potential, the rational design of CPPs remains a challenging task that often requires extensive experimental efforts and iterations. In this study, we introduce an innovative approach for the de novo design of CPPs, leveraging the strengths of machine learning (ML) and optimization algorithms. Our strategy, named LightCPPgen, integrates a LightGBM-based predictive model with a genetic algorithm (GA), enabling the systematic generation and optimization of CPP sequences. At the core of our methodology is the development of an accurate, efficient, and interpretable predictive model, which utilizes 20 explainable features to shed light on the critical factors influencing CPP translocation capacity. The CPP predictive model works synergistically with an optimization algorithm, which is tuned to enhance computational efficiency while maintaining optimization performance. The GA solutions specifically target the candidate sequences' penetrability score, while trying to maximize similarity with the original non-penetrating peptide in order to retain its original biological and physicochemical properties. By prioritizing the synthesis of only the most promising CPP candidates, LightCPPgen can drastically reduce the time and cost associated with wet lab experiments. In summary, our research makes a substantial contribution to the field of CPP design, offering a robust framework that combines ML and optimization techniques to facilitate the rational design of penetrating peptides, by enhancing the explainability and interpretability of the design process.

# 1 Introduction

In the modern era of nanotechnology and medicine, macromolecular entities such as peptides, proteins, and nucleic acids represent a promising prospect to overcome the limitations of conventional therapeutics[1]. However, the presence of cell and tissue barriers, together with the low transmembrane permeability of macromolecules, often hampers systemic drug distribution. Within this context, several peptides with membrane penetrating function[2–4] could support the intracellular delivery of hydrophilic macromolecules to eukaryotic cells. The ability of these small peptides to efficiently translocate into cells stimulated extensive structure-function analyses to identify the minimal amino acid sequence responsible for the translocation propensity[5,6]. A plethora of small peptides with membrane translocation capacities have been later identified[7–10]. They are termed cell-penetrating peptides (CPPs), which are short peptides that can successfully deliver proteins, peptides, siRNAs, and DNA across the lipid bilayer of the cell membrane into different cell types[11]. CPPs have demonstrated their usefulness in several fields such as cancer treatment[12–14], drug delivery [15–18], *in vivo* imaging and diagnostic [19–22], gene therapy[23–27], and radiotherapy[28–30]. CPPs can transport various classes of anticancer molecules with anticancer properties into cells, such as proteins, plasmid DNA, and oligonucleotides, and have often demonstrated selectivity for cancer cells over healthy cells [31–33]. This selectivity can also be employed to deliver imaging agents to tumour cells [19,31,34]. Predicting peptide penetration into biological membranes using computational methods has proven to be both an efficient and cost-effective approach for screening expansive chemical libraries[35–42].

Out of the numerous techniques used to enhance this predictive performance, one promising methodology has been the integration of machine learning (ML) models, pairing them with sequence-derived descriptors. Initial breakthroughs in the field include the CPPred-RF[38], which utilized the random forest (RF) algorithm integrated with several sequence-based descriptors. Soon after, Qiang et al. took a similar approach and developed the CPPred-FL, leveraging multiple RF models trained on various attributes, ranging from amino acid composition to specific position information[43]. Fu et al. proposed a unique support vector machine (SVM) algorithm, which predicts CPPs based entirely on the amino acid composition[44]. Subsequent advancements led to the creation of the KELM-CPPpred, a pioneering ML framework[36] that also relied heavily on amino acid composition. The combination of both sequence and structure-based descriptors resulted in improved accuracy in the CPP prediction. For instance, Manavalan et al. presented a framework that successfully integrated multiple techniques, including RF, SVM, extremely randomized trees (ERT), and k-nearest neighbour (K-NN) algorithms[35]. Drawing from a diverse array of data, such as amino acid composition and physicochemical properties, the study was able to distinguish between CPPs and non-CPPs more accurately. Furthermore, CellPPD-Mod[45] was then introduced as a powerful computational tool that uses the RF method to discriminate between CPPs and non-CPPs up to 25 residues-based end, molecular descriptors, and molecular fingerprints. Recently, an ML-based framework BChemRF-CPPred[46] combined different techniques such as ANN, SVM, and Gaussian classifier. Another development by Manavalan et al., called MLCPP2.0[47], encompassed a stacking ensemble of models to augment the prediction of CPP uptake efficiency. Recently, Zhang et al. proposed a Siamese neural network augmented by a contrastive learning optimization module for the discrimination between CPPs and non-CPPs[48]. This novel integrated approach has demonstrated exceptional performance and represents a reliable method for differentiating CPPs from non-CPPs. In the same vein, the SVM-based in silico tool CellPPD[49] stands as a practical application allowing users to generate all possible mutations and predict cell penetration.

Despite the recent advancement in this field, the importance of interpretable and explainable ML models cannot be overstated. These models not only offer pivotal insights into the performance of CPPs but also elucidate the underlying patterns and critical factors affecting their penetration abilities. Their ability to explain their predictions in understandable terms goes a long way in promoting the understanding of CPPs, ultimately paving the way for the development of better drug delivery systems. In the present study, we present LightCPPgen, an interpretable computational pipeline for virtual screening and generation of CPPs. In order to achieve interpretability and explainability, the machine learning model used in this study has been designed to provide clear and tangible rationales for its predictions, thereby enhancing confidence in its use and the resulting findings. The generative model is able to optimize non-penetrating peptides based on the solution of a constrained optimization problem with a genetic algorithm, where the cost function maximizes the similarity between the generated sequence and the input one. The constraint is applied such that the generated sequence is predicted to be penetrating by the ML model, but the underlying reasons for this prediction, as interpreted and explained by the model, are also provided.

# 2 Material and Methods

## 2.1 Dataset description

In this study, we employed the Layer1 training and testing datasets of MLCPP2.0[47]. The training dataset was derived from six existing methods, namely C2Pred[39], CellPPD[49], CPPred-RF[38], KELM-CPPpred[36], MLCPP[35], and BChemRF-CPPred[46]. Initially, positive and negative samples (CPPs and non-CPPs) were independently grouped. Subsequently, the CD-HIT[50] algorithm was applied to the positive samples, excluding CPPs that shared more than 85.0% sequence identity with other CPPs. This refinement process resulted in a set of 573 CPPs. Analogously, non-CPPs were processed, excluding sequences that shared >85.0% sequence identity with other non-CPPs or CPPs. To achieve a balanced training set, 573 non-CPPs were randomly selected. The positive samples of the test dataset were collected from the CPP entries of the CellPPD[49], KELM-CPPpred[36], MLCPP[35], and BChemRF-CPPred[46] independent datasets, the CPPsite 2.0[51], and Basith et al. CPP methods evaluation[52]. To ensure dissimilarity with the training dataset, sequences sharing high similarity were excluded using a CD-HIT[50] cutoff of 90.0%, yielding a total of 157 CPPs. Instead, the negative samples for the test dataset were collected from existing methods, excluding sequences with more than 70% sequence identity with the training samples, resulting in 2184 non-CPPs.

## 2.2 Methods

### 2.2.1 Feature engineering

In line with established literature[46,47,53], our investigation focuses on two principal categories of molecular descriptors. The first category encompasses descriptors based on the molecular structure, which relates to the physicochemical properties of the peptide. These are primarily derived from the aggregate characteristics of its constituent amino acids. Utilizing the RDKit library[54], we computed various structure-based descriptors including molecular weight (MW), the number of rotatable bonds (NRB), topological polar surface area (tPSA), fraction of sp3-hybridized carbon atoms (Fsp3), octanol-water partition coefficient (cLogP), number of aromatic rings (NAR), number of hydrogen bond donors (HBD) and acceptors (HBA), number of primary amino groups (NPA), number of guanidinium groups (NG), and the net charge (NetC). Additionally, we calculated the isoelectric point (IsoP) and hydrophobicity using the Python Peptides library, and aromaticity with the Biopython library[55]. Finally, we computed the length of the peptide sequence.

The second category includes sequence-based descriptors, directly derived from the peptide's primary amino acid sequence. For these, we employed the iFeatureOmega library[56] to calculate a comprehensive set of descriptors: amino acid composition (AAC), enhanced amino acid composition (EAAC), composition of k-spaced amino acid pairs (CKSAAP), di-peptide composition (DPC), tri-peptide composition (TPC), dipeptide deviation from expected mean (DDE), grouped amino acid composition (GAAC), enhanced GAAC (EGAAC), composition of k-spaced amino acid group pairs (CKSAAGP), grouped di-peptide composition (GDPC), grouped tri-peptide composition (GTPC), Moran correlation, Geary correlation, normalized Moreau-Broto autocorrelation (NMBroto), the composition, transition, and distribution set of features (CTDC, CTDT, CTDD), conjoint triad (CTriad), k-spaced conjoint triad (KSCTriad), sequence-order-coupling number (SOCNumber), quasi-sequence order (QSOrder), pseudo-amino acid composition (PAAC), amphiphilic pseudo-amino acid composition (APAAC), adaptive skip dipeptide composition (ASDC), auto covariance (AC), cross covariance (CC), auto-cross covariance (ACC), AAindex (AAINDEX), BLOSUM62, and Z-scale (ZSCALE).

## 2.2.2 Modeling approach: LightGBM

LightGBM[57], which stands for Light Gradient Boosting Machine, is an advanced implementation of the gradient boosting framework. This machine learning (ML) algorithm is highly praised for its efficiency and performance, especially in scenarios involving large datasets and the need for fast computation.

The fundamental concept of LightGBM, as with other gradient boosting methods, is to build an ensemble of weak prediction models, which are in most cases simple decision trees, sequentially. Each tree in the sequence attempts to correct the residuals or errors of the preceding trees. The final predictive model is an aggregation of these weaker decision trees. The model's output for a given input $x$ is given by:

$$F(x) = \sum_{t=1}^{N_T} \beta_t f_t(x)$$

Where $N_T$ represents the total number of trees, $f_t(x)$ is the output of the $t$-th decision tree and $\beta_t$ is the corresponding weight, often set to $\beta_t = 1/N_T M$.

In each iteration of training, the model focuses on correcting its previous mistakes. At each round, the negative gradient of the loss function with respect to the model's predictions is computed, which is called pseudo-residuals, and quantifies the direction and magnitude of the error. This gradient information is then used to guide the construction of the next tree, specifically aiming to reduce the residual error. The update formula can be expressed as:

$$F_{t+1}(x) = F_t(x) + \alpha f_{t+1}(x).$$

Here, $F_{t+1}(x)$ is the updated model after iteration $t+1$, $F_t(x)$ is the model from the previous iteration $t$, $f_{t+1}(x)$ is the newly added tree, and $\alpha$ is the learning rate, which controls how strongly each new tree influences the final model.

LightGBM introduces two innovative approaches to improve efficiency: Gradient-based One-Side Sampling (GOSS) and Exclusive Feature Bundling (EFB). GOSS improves the training process by prioritizing training instances that have larger gradients, meaning it focuses more on the harder-to-predict instances. This approach reduces the number of data instances needed in each iteration without significantly sacrificing accuracy. On the other hand, EFB efficiently manages sparse data by bundling together exclusive features, which are those features that are rarely non-zero at the same time. This reduces the number of features to consider, thus improving computational efficiency. These techniques allow LightGBM to handle large volumes of data with higher speed and lower memory usage compared to traditional gradient boosting methods. Moreover, LightGBM supports categorical features natively and is capable of handling missing data, making it a versatile tool in various machine learning tasks, from classification to regression. Despite its efficiency, it still maintains a high level of accuracy, making it a popular choice among data scientists and researchers.

### 2.2.3 Cost-sensitive learning

Cost-sensitive learning is a paradigm in ML where different types of prediction errors incur different costs[58]. Unlike the standard learning approach that treats all errors equally, cost-sensitive learning aims to minimize a cost function that reflects the real-world impact of the severity of different types of errors. This approach is particularly relevant when the consequences of different types of errors are not symmetrical. That is, the cost of classifying a positive example as negative can be very different from the cost of classifying a negative example as positive.

In the context of this work, which aims to suggest new candidate CPPs, we selected cost-sensitive learning to minimize false positives. Balancing the reduction of false positives with the risk of increasing false negatives is the key to ensure that the model remains effective in identifying true CPPs without missing potential discoveries. In order to address this point, a class weighting mechanism is incorporated into the cost function to control how the model learns during training. For instance, in the case of binary classification tasks where the log-likelihood function is used as the cost function:

$$L = \sum_{i=1}^{n} \left[ y^{(i)} \ln \left( P\left(y^{(i)}\right) \right) + \left(1 - y^{(i)}\right) \ln \left(1 - P\left(y^{(i)}\right)\right) \right]$$

where $P\left(y^{(i)}\right)$ denotes the predicted probability that $y^{(i)}$ equals 1, the cost function can be modified by introducing asymmetric costs $C_{FP}$ and $C_{FN}$ for false positives and false negatives, respectively:

$$L = \sum_{i=1}^{n} \left[ C_{FP} y^{(i)} \ln \left( P\left(y^{(i)}\right) \right) + C_{FN} \left(1 - y^{(i)}\right) \ln \left(1 - P\left(y^{(i)}\right)\right) \right].$$

In LightGBM, the modification of the cost function can be achieved by tuning a hyperparameter called `scale_pos_weight.` This hyperparameter, typically used to adjust for class imbalance by giving greater weight to the positive class, can be set to a value less than the default value of 1.0 to prioritize reducing false positives in binary classification tasks. This unconventional use inversely adjusts the model's focus, making it more conservative in predicting positives, thus leading to fewer false positives.

### 2.2.4 Model interpretation

In many scientific disciplines, the task of predicting a response variable from a set of predictor variables is crucial. While the primary goal of ML is often to maximize the accuracy of out-of-sample predictions, the identification of the most relevant predictor variables can hold equal importance. This is key to deepening our understanding of the underlying mechanisms or processes that drive these predictions. Additionally, it assists in making informed decisions in various practical applications. Due to these implications, the study and development of model interpretation techniques have been gaining considerable attention and popularity in recent years. These techniques not only aid in model transparency but also ensure that the findings and decisions based on ML models are reliable and justifiable.

## Global feature attribution methods

In ML, understanding the relative importance of input predictor variables is crucial. Typically, only a few variables influence the response (relevant variables), while many others may have minimal or no effect (weakly relevant or irrelevant variables). Quantifying the contribution of each input variable in predicting the response is called feature importance attribution.

In most implementations of tree-based ensemble models, such as Random Forest (RF) and Gradient Boosting (GB), feature importance is typically computed using the Mean Decrease in Impurity (MDI) method proposed by Breiman[59]. MDI is a method specifically tailored for tree-based models. A decision tree is a predictive model organized as a tree structure T, mapping input features $x \in \mathcal{X}$ to output targets $y \in \mathcal{Y}$. Typically, the input space $\mathcal{X}$ is represented as a $P$-dimensional real space $\mathbb{R}^P$, with $P$ denoting the number of input features. In classification tasks, $\mathcal{Y}$ comprises a set of discrete labels $\{1, 2, \dots, C\}$, where $C$ indicates the number of distinct, mutually exclusive classes. The decision tree for classification is known as a classification tree. Noteworthy, in binary classification where $C = 2$, the typical assumption is $y \in \{0,1\}$.

The decision tree uses the root node to represent the entire input space $\mathcal{X}$. Each internal node $t$ corresponds to a specific subset of $\mathcal{X}$, and the branches from these nodes result from binary decisions or splits $s_t = (x_k < c)$, further dividing the portion of input space associated with node $t$ into two child nodes $t_L$ and $t_R$. Nodes without children, situated at the ends of the tree, are known as terminal or leaf nodes. A test instance $x$ is classified by traversing from the root down to a leaf node, where in classification tasks, each leaf node assigns a prediction $\hat{y}$ based on the majority class of the training samples that end there. This structure enables the tree to adaptively map various regions of $\mathcal{X}$ to the appropriate output predictions, mirroring the training data's underlying distribution.

During the training phase, the tree structure is developed starting from a dataset encapsulated at the root node. Nodes are recursively created via a greedy procedure that groups samples with identical labels values. At each node $t$, the algorithm selects the optimal split $s_t = s^*$ that maximizes the reduction in impurity:

$$\Delta i(s_t) = i(t) - w_{t_L} i(t_L) - w_{t_R} i(t_R)$$

Here, $i(t)$ represents an impurity measure (e.g., the Gini index, or the Shannon Entropy for classification), and $w_{t_L} = N_{t_L}/N_t$ and $w_{t_R} = N_{t_R}/N_t$ denote the proportions of samples at node $t$ moving to $t_L$ and $t_R$ respectively. The effectiveness of each split, quantified by $\Delta i(s_t)$, signifies how well it purifies the node. Tree construction typically ceases when the nodes are pure or when no further informative splits can be made.

Practical implementations may also include constraints like maximum tree depth or minimum sample count per leaf to prevent overfitting. In an ensemble approach, multiple decision trees ($N_T$ in total) are built, and the importance $I_j$ of each input feature $x_j$ in predicting the target is evaluated by averaging the weighted decreases in impurity from all nodes where $x_j$ influences the split, across all trees:

$$I^{MDI}_j = \frac{1}{N_T} \sum_T \sum_{t \in T : v(s_t) = x_j} w_t \Delta i(s_t)$$

where $w_t$ is the proportion $N_t/N$ of samples reaching node $t$, and $v(s_t)$ indicates the feature used in the split. This measure, known as Mean Decrease in Impurity (MDI), reflects the feature's overall impact on model accuracy, highlighting the importance of features used in early splits due to their influence on a larger portion of the input space. This underscores their significant role in the decision-making process of the model.

Unfortunately, feature importance methods based on impurity suffer from a significant drawback as they tend to overstate the importance of features with high cardinality (typically numerical features), meaning variables that possess many unique values. For instance, this contrasts with binary or categorical features that have a small finite number of possible categories. This bias towards continuous features with high cardinality stems from their inherent ability to provide more potential splitting points. These points, by chance, could lead to a substantial decrease in impurity when a node is split. The most striking consequence of such a phenomenon is that even randomly generated features, which have no actual association with the target variable but possess high cardinality, can be erroneously classified as highly important. This misclassification occurs if the model is sufficiently flexible to overfit these variables, leading to serious inconsistencies in assessing the actual contribution of features. Unfortunately, these inconsistencies stem from the way feature importance is calculated using impurity-based methods. Specifically, this importance is assessed during the training phase and is thus biased towards the training data. As a result, it does not accurately reflect the utility of features in making generalizable predictions on new data. Additionally, impurity-based attributions also fail to effectively handle features that are strongly correlated or, more generally, highly dependent. In particular, they tend to underestimate the importance of multicollinear variables that impact the prediction of the target variable. This underestimation occurs because these variables provide access to equivalent information through different channels, causing a dilution effect on the estimated importance, which gets divided among the correlated variables. Despite the issues, MDI feature importance can still be used to identify variables deemed irrelevant by the model, namely those never selected for a split. These variables do not influence the final prediction, resulting in an MDI value of zero.

## Local feature attribution methods

The feature attribution methods discussed in the previous section are known as global feature attribution methods. They describe the overall behavior of an ML model, focusing on providing an aggregated and holistic understanding of how features generally influence a model's predictions. In contrast, the local feature attribution methods aim to explain individual predictions made by a ML model. These methods offer specific insights into why a model made a particular prediction for a single instance. A notable example of local feature attribution methods involves the use of Shapley values, originating from cooperative game theory.

In cooperative games, all players, denoted as $P = \{1, \dots, p\}$ and collectively referred to as the grand coalition, can form any subset $S \subseteq P$, each representing a potential coalition. The value of a coalition, $v(S)$, quantifies the collective payoff that its members can achieve together. This value is defined by a characteristic function $v : 2^P \mapsto \mathbb{R}$, mapping each coalition to its respective payoff, with the empty coalition necessarily having a zero value ($v(\emptyset) = 0$). Cooperative game theory primarily seeks to fairly distribute the grand coalition's total achievable payoff, $v(P)$, among players based on their contributions. This distribution is represented by the individual payoff $\phi_j(v)$ for player $j$, or simply $\phi_j$. The Shapley value, introduced by Lloyd Shapley in 1951, offers a method to calculate these payoffs, ensuring each player receives a share equal to their average marginal contribution across all possible coalitions. The marginal contribution of player $j$ to a coalition $S$ is the difference in value when player $j$ is added, expressed as $v(S \cup \{j\}) - v(S)$. This measure reflects the added value player $j$ brings to the coalition.

Mathematically, the Shapley value is calculated as:

$$\phi_j = \sum_{S \subseteq P \setminus \{j\}} \frac{|S|! \, (|P| - |S| - 1)!}{|P|!} \big( v(S \cup \{j\}) - v(S) \big)$$

where the summation considers all possible coalitions excluding players. Each contribution is weighted by the number of permutations of players in S and adjusted by total permutations $|P|!$, ensuring all contributions are proportionally represented.

The Shapley value is the only attribution strategy satisfying the four desirable properties of:

- *Efficiency*: the sum of individual payoffs equals the total value achievable by the grand coalition:

$$\sum_{j=1}^{P} \phi_j = v(P)$$

- *Symmetry*: players who contribute equally to all coalitions receive identical payoffs:

$$v(S \cup \{j\}) = v(S \cup \{k\}) \text{ for all } S \implies \phi_j = \phi_k$$

- *Null player*: any player who does not enhance any coalition's value receives no payoff:

$$v(S \cup \{j\}) = v(S) \text{ for all } S \Rightarrow \phi_j = 0$$

- *Additivity*: if two games with value functions $v$ and $w$ are combined, the resulting payoffs equal the sum of payoffs from each game separately:

$$\phi_j(v + w) = \phi_j(v) + \phi_j(w)$$

The application of Shapley values to ML for explaining model predictions was first proposed by Erik Štrumbelj and Igor Kononenko in 2010[60]. In this context, a cooperative game is analogous to forming a prediction $\hat{y} = f(x_1, \ldots, x_p) = f(x)$, where $f$ is the model or the prediction function and $x$ represents an individual observation. Each feature $x_j$ is considered a player that interacts with others to generate a prediction. The primary challenge is predicting with the model $f$ trained using all features but leveraging only a subset of them. Given the complete feature set $X = \{x_1, \ldots, x_p\}$, we define a coalition of features $S \subseteq P$, which translates to a feature subset $X_S \subseteq X$. The remaining features outside the coalition form the complementary feature subset $X_C = X \setminus X_S$, and the goal is to compute the reduced prediction $f_S(x_S)$, using only the features in $S$.

One solution is to marginalize the features not in the coalition by using the partial dependence function, defined as:

$$f_S(x_S) = pd_{X_S}(x_S) = \mathbb{E}_{X_C}[f(x_S, X_C)] = \int f(x_S, x_C)p(x_C)dx_C$$

where $f(x_S, x_C)$ is the prediction for observation $x$, which values are defined by the tuples $x_S$ for the features in $X_S$ and $x_C$ for the features in $X_C$. To determine the value of a coalition $S \subseteq P$ for an observation $x$, we calculate:

$$v(S) = f_S(x_S) - \mathbb{E}_X[f(X)]$$

Subtracting the average prediction $\mathbb{E}_X[f(X)]$ ensures that the empty coalition has a value of zero. The marginal contribution of the $j$-th feature is given by:

$$v(S \cup \{j\}) - v(S) = f_{S \cup \{j\}}(x_{S \cup \{j\}}) - f_S(x_S)$$

This difference reflects the prediction change when the $j$-th feature is added to coalition $S$. Thus, the Shapley value $\phi_j$ for the $j$-th feature is computed for the observation $x$:

$$\phi_j(x) = \sum_{S \subseteq P \setminus \{j\}} \frac{|S|! \, (|P| - |S| - 1)!}{|P|!} \left( f_{S \cup \{j\}}\left(x_{S \cup \{j\}}\right) - f_S(x_S) \right)$$

This is the weighted average of all possible coalitions excluding $j$, reflecting the marginal contributions of the $j$-th feature to the model prediction.

The fair attribution axioms, in the ML context, have new useful interpretations:

- *Efficiency*: the Shapley values sum up to the final prediction $f(x)$ minus the average prediction $\mathbb{E}_X[f(X)]$:

$$\sum_{j=1}^{p} \phi_j(x) = f(x) - \mathbb{E}_X[f(X)]$$

- *Symmetry*: two features $x_j$ and $x_k$ that contribute equally to the prediction receive the same payoff. This ensures a fair division of feature interactions.
- *Null Player*: a feature that does not influence the prediction receives a zero Shapley value. This applies, for example, if a feature isn't selected by a model like LASSO or tree-based models.
- *Additivity*: When combining two predictions, a feature's total contribution is the sum of contributions to the individual predictions. This axiom is crucial for ensemble models like RF or GB, ensuring aggregation of Shapley values calculated for different base learners.

The advantages of Shapley's values include their strong theoretical foundation in game theory and their ease of interpretation. However, computing them exactly is challenging due to two reasons:

1. Feature distribution: the value function requires integration over feature distributions, but real-world data often only provides a finite dataset.
2. Exponential Complexity: calculating all feature combinations is infeasible due to the exponential growth of possible coalitions ($2^P$).

Monte Carlo integration can approximate the value function using the available dataset $\mathcal{D} = \{x^{(i)}\}_{i=1}^{n}$:

$$v(S) \approx \frac{1}{n} \sum_{i=1}^{n} \left( f\left(x_S, x_C^{(i)}\right) - f(x^{(i)}) \right)$$

Substituting this approximation into the Shapley Value formula provides an estimate:

$$\phi_j(x) \approx \frac{1}{n} \sum_{S \subseteq P \setminus \{j\}} \frac{|S|! \, (|P| - |S| - 1)!}{|P|!} \sum_{i=1}^{n} \left( f\left(x_{S \cup \{j\}}, x_{C \setminus \{j\}}^{(i)}\right) - f\left(x_S, x_C^{(i)}\right) \right)$$

Since the number of coalitions grows exponentially, it's vital to employ estimation techniques that avoid exhaustive computation. Shapley Additive explanations (SHAP), implemented in the Python SHAP library, provide a practical solution. SHAP optimizes calculations for ensemble models, reducing computational complexity from $O(TL2^P)$ to $O(TLD^2)$, where $T$ is the number of trees, $L$ is the maximum number of leaves, and $L$ is the maximum depth.

Shapley Values represent local feature attributions, reflecting each feature's influence on the prediction at the level of an individual observation $x$. However, aggregating Shapley Values across a dataset provides a global view of feature importance via a matrix $\Phi$, where each row corresponds to an observation, and each column to a feature:

$$\Phi = \left( \phi_j^{(i)} \right)_{i=1,\ldots,n \, , \, j=1,\ldots,P}$$

Feature importance can be derived by averaging the absolute Shapley Values per feature:

$$I^{SHAP}{}_j = \frac{1}{n} \sum_{i=1}^{n} \left| \phi_j^{(i)} \right|$$

Like many other interpretation methods, the Shapley value approach encounters challenges when dealing with correlated features [61]. To simulate the absence of a feature from a coalition, the method involves marginalizing that feature, typically by sampling from the feature's marginal distribution. This approach works well for independent features. However, when features are dependent, this sampling process may introduce unrealistic feature values for the given instance. One possible solution is to use the `PartitionExplainer` from the SHAP library, which requires as argument a hierarchical clustering of the input features based on correlation. The `PartitionExplainer` computes Shapley values recursively through the hierarchy of features that defines feature coalitions, considering groups feature together and allocating credit based on how one group would perform as a whole. This means that if the clustering provided to the `PartitionExplainer` groups these correlated features together, the method effectively manages feature correlations. Essentially, the total credit given to a group of closely related features remains consistent and does not fluctuate based on changes in their correlation during the explanation process. This ensures that the explanation reflects their combined influence accurately, without being affected by any alterations to their interdependencies.

## 2.2.5 Feature selection

In this study, we developed a comprehensive set of 13,833 descriptors to thoroughly explore the feature space, aiming to identify distinctive factors between penetrating and non-penetrating peptides. This extensive feature set allows for a deep and potentially novel exploration of correlations. However, it also presents several challenges:

1. *Computational Load*: The calculation of a large number of features, coupled with model training and prediction, becomes computationally demanding in terms of speed and memory usage.
2. *Risk of Overfitting*: With such a vast feature set, there's an increased likelihood of the model capturing noise rather than the underlying pattern, leading to poor generalization to new data.
3. *Complex Interpretability*: A high number of features can obscure the understanding of what drives the model's predictions, making interpretation less straightforward.

Feature selection emerges as a crucial step in this research to solve the above-mentioned challenges.

Given a set of $p$ features indexed by $P = 1, \ldots, p$, the problem of feature selection consists in identifying the optimal subset $S^* \subseteq P$ that optimizes an objective function $\Psi : \Omega \mapsto \mathbb{R}$, expressed as:

$$S^* \in \text{argmin}_{S \in \Omega} \Psi(S)$$

Here, the search space $\Omega$, includes all possible feature subsets $S \subseteq P$, in other words, is the power set of $P$ ($\Omega = 2^P$). For convenience, $\Omega$ can be represented as bit vectors of length $p$, where each bit at position $j$ indicates the inclusion of the $j$-th feature in subset $S$. This setup frames the feature selection task as a combinatorial optimization problem with a search space growing exponentially with $p$. In general, this problem is NP-hard and cannot be solved efficiently.

The objective function $\Psi$ typically estimates the generalization error, and its choice determines whether the feature selection algorithm is a wrapper, filter, or embedded method.

Wrapper methods, conceptualized by Ron Kohavi in 1997[62], arise from the observation that optimal feature sets vary across different learning algorithms. These methods aim to directly solve the optimization problem by training a predictive model for each subset of features $S$ and evaluating the objective function $\Psi(S)$ using in-sample metrics like AIC/BIC or out-of-sample metrics like cross-validation performance. Although wrapper methods can be computationally intensive due to the need to train models for each feature subset, they provide the optimal feature subset for the specific model configuration and evaluation criterion used. Efficient search strategies, such as sequential feature selection heuristics like greedy forward and backward search, can mitigate the need for exhaustive search.

Embedded methods, like LASSO or tree-based approaches, integrate feature selection within the model training phase, promoting sparse solutions, or implicitly selecting features during tree construction. These methods balance feature relevance and model complexity directly through the learning algorithm.

In our work, we use a hybrid feature selection strategy that merges MDI feature selection (an embedded method) and greedy forward search (a wrapper method). Initially, a LightGBM model is trained using the entire feature set $P$, and the MDI feature importance $I_j$ computed internally by the model for each feature $j \in P$ is used to prune the feature set to $\bar{P} = \{j \in P : I_j > 0\}$. This preprocessing step significantly reduces the feature pool, enabling a more efficient subsequent application of the greedy forward search.

Finally, from the reduced feature set $\bar{P}$, the greedy forward search method is used to find the final feature subset $S^*$ as follows:

1. Start with an empty feature set $S = \emptyset$.
2. Repeat {
   a. Generate candidate subsets $S_j = S \cup \{j\}$ for each $j \notin S$.
   b. For each $j \notin S$ train a model using $S_j$ and estimate its generalization error.
   c. Update $S$ to the best $S_j$ found in step b. }
3. Output the best feature subset $S^*$ evaluated during the entire search procedure, or until a stopping criterion is met (e.g., maximum number of features to select).

## 2.2.6 Validation strategy and evaluation criteria

To assess the out-of-sample performance of the model during hyperparameter tuning and feature subset evaluation, we employed stratified $k$-fold cross-validation. This approach partitions the training set into $k$ smaller subsets (or folds) through stratified sampling, ensuring that the proportion of samples from each class mirrors that of the entire training set. In this process, each fold serves as a validation set once, while the model is trained on the combined data from the remaining $k-1$ folds. The overall performance from the cross-validation is then calculated as the mean of the individual performances across all $k$ folds. For performance evaluation of the finalized model and its comparison with models cited in the literature, we utilized the independent dataset from MLCPP 2.0 as our test set.

This study uses both threshold-dependent and threshold-independent metrics to evaluate binary classification performance. The threshold-dependent metrics employed are Accuracy (ACC), Sensitivity (SN), Specificity (SP), and the Matthews correlation coefficient (MCC), which are defined as follows:

$$\text{ACC} = \frac{\text{TP} + \text{TN}}{\text{TP} + \text{FP} + \text{TN} + \text{FN}} \times 100,$$

$$\text{SN} = \frac{\text{TP}}{\text{TP} + \text{FN}} \times 100,$$

$$\text{SP} = \frac{\text{TN}}{\text{TN} + \text{FP}} \times 100,$$

$$\text{MCC} = \frac{\text{TP} \times \text{TN} - \text{FP} \times \text{FN}}{\sqrt{(\text{TP} + \text{FP})(\text{TP} + \text{FN})(\text{TN} + \text{FP})(\text{TN} + \text{FN})}}.$$

Here, TP (True Positives) and TN (True Negatives) represent the counts of correctly predicted CPP and non-CPP samples, respectively, while FP (False Positives) and FN (False Negatives) denote the counts of incorrectly predicted CPP and non-CPP samples, respectively. ACC quantifies the overall percentage of correct predictions made by the model. SN, or the true positive rate, measures the proportion of actual CPPs correctly identified as such, whereas SP, or the true negative rate, reflects the proportion of non-CPPs accurately identified. MCC offers a balanced metric, particularly useful when class sizes are unbalanced, and ranges from $-1$ (total disagreement) to $+1$ (perfect prediction), with $0$ indicating no better than random prediction. Additionally, the Area Under the Receiver Operating Characteristic Curve (AUC) serves as a threshold-independent metric to evaluate model performance.

### 2.2.7 Model reliability

Among different possibilities to assess the reliability of the model, we adopted the Local Outlier Factor (LOF)[63] unsupervised learning algorithm for novelty detection, as described below.

Let $\mathcal{D}_T = \left\{ \left( x^{(i)}, y^{(i)} \right) \right\}_{i=1}^N$ be the $N$-dimensional set used for training, with $x^{(i)}$ and $y^{(i)}$ being the feature and label, respectively, associated to the $i$-th sample. Let $x^{(N+1)}$ be the feature associated with a new sample. We assess the reliability of the model by measuring the level of novelty of $x^{(N+1)}$ with respect to features used for training, collected in the set $\mathcal{X}_T = \left\{ x^{(i)} \right\}_{i=1}^N$. In particular, the reliability of the model associated with the new feature $x^{(N+1)}$ decreases as the deviation of $x^{(N+1)}$ with respect to the training dataset $\mathcal{X}_T$ increases.

The LOF algorithm measures the local density deviation of a given data point $x^{(N+1)}$ with respect to its neighbors, by comparing the density of areas surrounding $x^{(N+1)}$ to the densities of the areas surrounding its neighbors. A high LOF value indicates that the sample is far from its neighbors (i.e., an outlier or a novel sample with respect to $\mathcal{X}_T$). Thus, it deviates from the training data, and this can suggest that the model's reliability is low for this sample. Conversely, a low LOF value suggests the sample is similar to its neighbors, suggesting high model reliability for this sample. The main steps of the LOF algorithms are reported below, assuming the features are already normalized:

- Let us define the augmented dataset $\tilde{\mathcal{X}} = \mathcal{X}_T \cup x^{(N+1)}$. For each pair of points $x, q \in \tilde{\mathcal{X}}$ compute the Euclidean distance between $x$ and $q$. Let us denote as $d(x, q)$ such a distance.
- For a positive integer k such that $0 < k \le N$ and each point $x \in \tilde{\mathcal{X}}$, compute the so-called k-distance (denoted as $\text{dist}_k(x)$), being the distance of $x$ with its k-th nearest point.
- For each point $x \in \tilde{\mathcal{X}}$, compute the k-nearest neighbors of $x$ (denoted as $N_k(x)$) defined as the set of samples of $\tilde{\mathcal{X}}$ (excluding $x$) with distance smaller than $\text{dist}_k(x)$. Basically, $N_k(x)$ includes all the points of $\tilde{\mathcal{X}}$ that lie in the circle centered at $x$ and of radius $\text{dist}_k(x)$. Denote as $|N_k(x)|$ the cardinality of $N_k(x)$, which can be larger or equal than k.
- For a pair of points $x, q \in \tilde{\mathcal{X}}$, define the reachability distance of $x$ from $q$ defined as

$$\text{rdist}_k(x, q) = max\{d(x, q), dist_k(q)\}$$

- For each point x $\in \widetilde{\mathcal{X}}$, compute the local reachability density of x defined as

$$\text{lrd}_k(x) = \frac{|N_k(x)|}{\sum_{q \in N_k(x)} rdist_k(x, q)}$$

The local reachability density is thus given by the inverse of the average reachability distance of the point $x$ from its neighbors. Low values of local reachability density thus mean that the point $x$ is "far" from its neighbors.

- For each point $x \in \widetilde{\mathcal{X}}$, the $lrd_k(x)$ is then compared with the local reachability density of its neighbors to compute the value of the local outlier factor for x, specifically:

$$\text{LOF}_k(x) = \frac{\sum_{q \in N_k(x)} lrd_k(q)}{|N_k(x)| lrd_k(x)}$$

Note that a value of $\text{LOF}_k(x)$ of approximately 1 indicates that the "local density" of the point $x$ is comparable to one of its neighbors. Thus, $x$ is not an outlier. On the other hand, values of $\text{LOF}_k(x)$ drastically larger than 1 indicate that $x$ is an outlier.

It is worth pointing out that if we are only interested in detecting if a new point $x^{(N+1)}$ is a novelty, there is no need to compute the LOF value for all points in the dataset $\widetilde{\mathcal{X}}$. Indeed, it is enough to compute the k-nearest neighbors of $x^{(N+1)}$, along with the local reachability density of its neighbors.

### 2.2.8 Optimal design of penetrating peptides

Starting from a non-penetrating peptide, we aim at generating a new penetrating amino acid sequence satisfying the following additional properties:

- it should be "close" to the original non-penetrating peptide, where closeness is measured according to the similarity score obtained from the global pairwise sequence alignment with the BLOSUM62 substitution matrix, using Biopython library[55].

- it should exhibit a low level of non-novelty relative to the dataset used for training the LightGBM model. In fact, the estimated LightGBM model serves as a surrogate during the design process, predicting the penetrating probability of a given peptide. Therefore, ensuring high reliability of the model is crucial during the design process.

According to the above considerations, the following performance metric is adopted as a criterion to guide the design, and thus used as a fitness function to be minimized by a genetic algorithm:

$$J(\theta) = w_1 \cdot d_{sim(\theta,\theta_0)} + w_2 \cdot \max\left(0, \left(1 - p(y = 1|\theta)\right)^2 - 0.2^2\right) + w_3 \cdot \max\left(0, \text{LOF}_{k(\theta)}^2 - 1.5^2\right)$$

In the definition of the fitness function $J(\theta)$:

- $\theta$ represents the vector of design parameters, specifically the ordered sequence of amino acids that characterize the peptide. Each element of this vector represents an amino acid, denoted by a Latin character in the one-letter code system (with an alphabet of 20 characters), making the design a combinatorial optimization problem. The dimension of the vector $\theta$ determines the length of the amino acid sequence, which is intended to be pre-specified by the user;

- $\theta_0$ is a given vector representing the amino acid sequence of the initial non-penetrating peptide;

- $d_{sim(\theta,\theta_0)}$ is the similarity distance between the candidate and the initial peptide;

- $p(y = 1|\theta)$ is the probability of a candidate peptide (characterized by the amino acid sequence $\theta$) to be penetrating;

- $LOF_{k(\theta)}$ is the value of the Local Outlier Factor for the candidate peptide.
- $w_1, w_2,$ and $w_3$ represents the weights by which it is possible to modify the impact of one or more factors in the fitness function $J(\theta)$. In our application they are all set to 1.

According to the above definition of the fitness function $J(\theta)$, the similarity distance $d_{sim(\theta, \theta_0)}$ between the candidate and initial peptide is minimized. Furthermore, polynomial barrier functions are used to penalize LOF novelty values larger than 1.5 and the probability of penetrability lower than 0.8.

A genetic optimization algorithm, whose main steps are summarized below, is used to minimize the fitness function $J(\theta)$ over the design parameters $\theta$:

1. **Initialization**: Create an initial population of $P_{(init)}$ potential solutions to the problem. These solutions are generated randomly and represent initial guesses of the peptides.

2. **Fitness Evaluation and Selection**: Evaluate the fitness $J(\theta)$ of each individual in the population and select $N_{\{sel\}}$ individuals with the lowest fitness value from the current population to serve as parents for the next generation. $N_{\{sel\}} = 100$ in our application.

3. **Crossover (Recombination)**: Perform crossover or recombination on pairs of selected parents to create offspring. This step involves combining the genetic information (namely, amino acids in our problem) of two parents to produce one or more offspring. Crossover is performed by randomly selecting, with uniform probability, the chromosome (amino acid) of one of the two parents as a chromosome for the child.

4. **Mutation**: Apply mutation to some of the offspring. Mutation introduces small random changes to amino acids in the sequences, thus maintaining genetic diversity in the population and promoting the exploration of new solutions. In our application, the probability of mutating a chromosome is 10%.

5. **Replacement**: Replace the old population with the new generation of individuals (offspring and some surviving parents). The population size remains constant. In our application, 10 parents with the lowest fitness survive from one iteration to another.

6. **Termination Criterion**: Repeat selection, crossover, mutation, and replacement steps for a predefined number of generations or until a termination criterion is met, such as reaching a target fitness or running for a specified time. In our application, a maximum of 50 iterations are performed, and the optimization stops when the best candidate satisfies the following criteria: $J(\theta) \leq 0.1$, $p(y = 1|\theta) \geq 0.95$, $LOF_k(\theta) \leq 1.3$.

7. **Result Extraction**: The individual with the best fitness in the final population represents the optimized solution to the problem.

A schematic depiction of the CPP design algorithm is shown in Figure 1. The code of the CPP design algorithm is freely available in the GitHub repository (https://github.com/gabribg88/LightCPPgen.git).

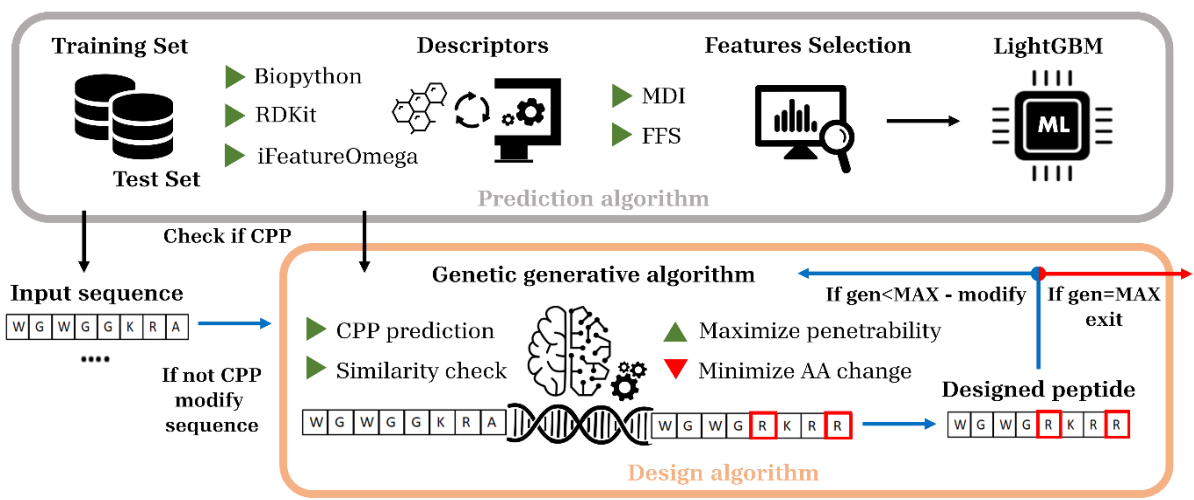

*Figure 1: Schematic representation of the cell penetrating peptides design process.*

# 3 Results

In this section, we show the results of the *de novo* CPP design employing a synergistic approach that integrates machine learning (ML) and genetic algorithms (GA). In detail, we present the performances of the lightGBM-based predictor (sections 3.1 and 3.2), the model explainability and interpretability (section 3.3), and the design algorithm (section 3.4). The evaluation of the design algorithm effectiveness involved the utilization of the non-CPP samples from the MLCPP2.0 indipendent dataset.

## 3.1 Modeling and feature selection results

In this study, we developed LightCPP, a LightGBM-based predictor for binary classification to distinguish between CPP sequences and non-CPP sequences. The numerical results presented in this section were obtained from testing the performance of the algorithm on Layer1 of the independent dataset from the public MLCPP 2.0 dataset. Training details are provided at the end of this section, following the results of the feature selection phase, which are discussed first.

As detailed in Section 2.2.5, initially, a LightGBM model was trained using the entire feature set comprising 13,833 variables. Based on the feature importance calculated using the MDI approach, all features deemed irrelevant by the model were discarded, reducing the feature set to 375 variables. This substantial reduction enabled the subsequent execution of the forward search within a reasonable time frame. To benchmark model performance before and after sequential feature selection, we recorded the performance of the model, referred to as LightCPP (375 features), which achieved metrics of 0.930, 96.2%, 69.0%, 98.1%, and 0.687 for AUC, ACC, SN, SP, and MCC, respectively.

In the subsequent phase, a greedy forward search was conducted, using MCC metric to evaluate feature subsets resulting in a selected subset of 20 features, described in detail in Section S1 - SI file. The final model, referred to as LightCPP (20 features), demonstrated performance metrics of 0.909, 96.0%, 69.0%, 97.9%, and 0.677 for AUC, ACC, SN, SP, and MCC, respectively. Compared to the intermediate model, there was a slight decrease in performance metrics by 2.26%, 0.21%, 0.0%, 0.20%, and 1.46% for AUC, ACC, SN, SP, and MCC, respectively. Despite the reduced number of features, the final model demonstrates a minor overall reduction in performance across most metrics, indicating the effectiveness of the feature selection algorithm.

Each training session was conducted on Layer1 of the MLCPP 2.0 training dataset using a stratified 10-fold cross-validation approach. The number of boosting iterations (`num_iterations`), corresponding to the number $N_T$ of trees in the final ensemble, was chosen to maximize the average AUC calculated across the 10 validation folds, with an early stopping round (`early_stopping_round`) set at 20. For all training sessions, the hyperparameter regulating the weight of labels with positive class (`scale_pos_weight`) was set to 0.1 to penalize false positives 10 times more than false negatives. The number of leaves (`num_leaves`) of the base learners was set to a minimum of 2 to minimize the inference times of the final model. Finally, we set the maximum number of features to select during the forward selection step at 20 as stopping criterion.

## 3.2 Comparison with the state-of-the-art predictors

To assess the generalization performance of LightCPP, we compared its metrics with those of various state-of-the-art models cited in the literature, specifically focusing on comparisons with models that utilize the independent dataset from MLCPP 2.0 as a test set, including C2Pred, BChemRF-CPPred, MCLPP, MLCPP 2.0 (Layer1), and SiameseCPP. In detail, these comparisons were made using Layer1 of the independent dataset from MLCPP 2.0, referencing the performance metrics listed in Table 2 from Zhang et al. 2023[48] and Table 10 from Manavalan et al.[47]. The numerical results are presented in Table 1. The best and second-best performances for each metric across all models are highlighted in bold and underlined, respectively.

*Table 1: Performance detail of our proposed LightCPP in comparison with C2Pred, BChemRF-CPPred, MLCPP, MLCPP2.0, and SiameseCPP on the Layer1 of the independent dataset from MLCPP 2.0.*

| Model | AUC | ACC (%) | SN (%) | SP (%) | MCC |
|---|---|---|---|---|---|
| C2Pred | 0.867 | 78.1 | 79.0 | 78.1 | 0.326 |
| BChemRF-CPPred | 0.914 | 89.3 | 74.5 | 90.3 | 0.467 |
| MLCCP | 0.920 | 89.2 | <u>80.9</u> | 89.8 | 0.497 |
| MLCPP2.0 (Layer1) | <u>0.928</u> | 93.4 | **84.7** | 94.0 | 0.624 |
| SiameseCPP | - | 95.9 | 62.4 | **98.3** | 0.652 |
| LightCPP (375 features) | **0.930** | **96.2** | 69.0 | <u>98.1</u> | **0.687** |
| LightCPP (20 features) | 0.909 | <u>96.0</u> | 69.0 | 97.9 | <u>0.677</u> |

The data in Table 1 indicate that no individual model consistently outperforms all others across all evaluation metrics. However, the models proposed in this manuscript exhibit the best performance in terms of AUC, ACC, and MCC (LightCPP (375 features)) and the second-best performance in terms of ACC, SP, and MCC (LightCPP (20 features)). The models that closely match the performance of our proposed models are MLCPP 2.0 (Layer1) and SiameseCPP, which deserve a more detailed comparative analysis.

When directly comparing the proposed models with the MLCPP 2.0 (Layer1) model, we see that in terms of AUC, all classifiers perform similarly with high values above 0.9, with the 20-feature model performing about 2% worse than the other two. In terms of ACC, the proposed models perform nearly 3% better than MLCPP 2.0 (Layer1). However, high ACC values might indicate a bias towards predicting the majority class (non-CPPs in this case), which is expected given the training strategy of the proposed models in term of penalizing false positive more than false negative. In terms of SP and SN, LightCPP predictors are substantially better at avoiding false positives while MLCPP 2.0 (Layer1) excels at identifying positive cases, thus countering false negatives. Finally, the proposed models also score higher in MCC, suggesting superior overall quality across all quadrants of the confusion matrix. Comparing the proposed models with SiameseCPP, we observe that in terms of ACC, the models are comparable; in terms of SN, the proposed models perform nearly 10% better, thus demonstrating a superior ability to correctly identify CPPs compared to SiameseCPP. In terms of SP, SiameseCPP shows a slightly higher score, but the difference is marginal. Finally, the MCC values suggest that LightCPP models are more balanced in their performance, reflecting better handling of both positive and negative classes.

## 3.3 Explainability of LightCPP

Understanding the importance of underlying features behind ML-based predictions is crucial for its effective application in CPP design. The 20 features governing the predictive performance of our developed LightCPP (20 features - Table 1), as detailed in Section S1 - SI file, shed light on crucial aspects of the model's predictive capabilities. The feature importance and impact, computed with SHAP feature importance on the indipendent set of MLCPP 2.0 are shown in Figure 2.

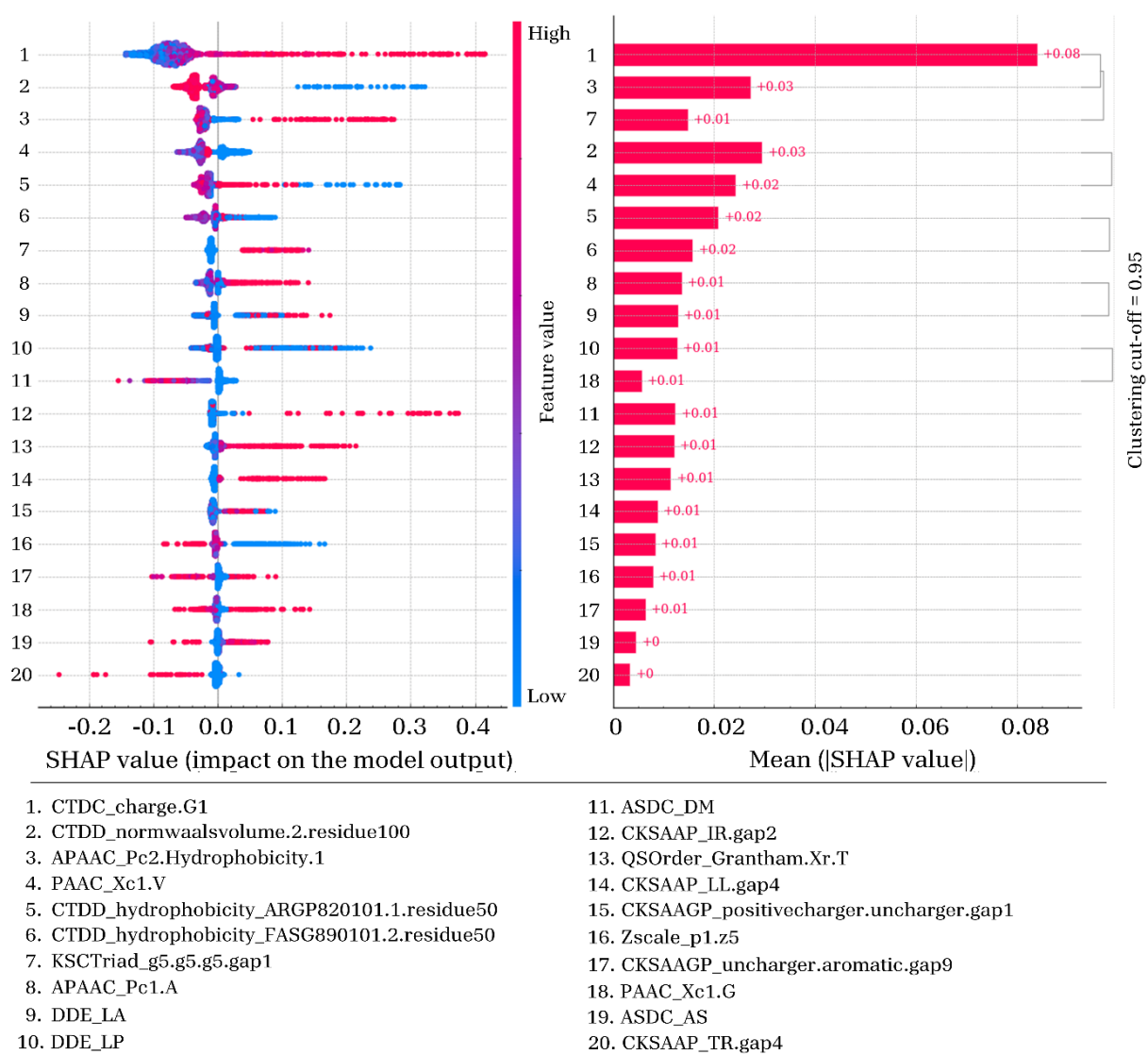

1. CTDC_charge.G1
2. CTDD_normwaalsvolume.2.residue100
3. APAAC_Pc2.Hydrophobicity.1
4. PAAC_Xc1.V
5. CTDD_hydrophobicity_ARGP820101.1.residue50
6. CTDD_hydrophobicity_FASG890101.2.residue50
7. KSCTriad_g5.g5.g5.gap1
8. APAAC_Pc1.A
9. DDE_LA
10. DDE_LP
11. ASDC_DM
12. CKSAAP_IR.gap2
13. QSOrder_Grantham.Xr.T
14. CKSAAP_LL.gap4
15. CKSAAGP_positivecharger.uncharger.gap1
16. Zscale_p1.z5
17. CKSAAGP_uncharger.aromatic.gap9
18. PAAC_Xc1.G
19. ASDC_AS
20. CKSAAP_TR.gap4

*Figure 2: Feature importance and impact as computed with SHAP on the indipendent set of MLCPP 2.0. The right bar plot ranks the variables by their average impact on model prediction. Clustering of features, as shown on the right, is based on computed feature dependence, as detailed in Section S2 – SI file. The left dot plot shows each data point with the signed contribution of each feature. Blue dots indicate low variable values, while red dots indicate high values.*

As expected, the CTDC charge feature (CTDC_charge.G1) demonstrated the importance of positively charged amino acids in predicting CPP behavior. High values of positive charge were consistently associated with the prediction of CPP, while low values were indicative of non-penetrating peptides. Moreover, the distribution along the peptide of the group of amino acids asparagine, valine, glutamic acid, glutamine, isoleucine, and leucine (CTDD_normwaalsvolume.2.residue100), emerged as another critical feature influencing CPP prediction. Low values of this feature lead to CPP classification, suggesting either a complete absence of these amino acids or their presence predominantly in the initial segment of the sequence. Contrariwise, high values indicate the presence of at least one of these residues close to the C-terminal tail of the sequence and results in the classification of non-CPP. The amphiphilic pseudo amino acid composition (APAAC), which reflects the balance of hydrophobicity and hydrophilicity along with the sequence-order correlation based on adjacent residues' hydrophobicity (APAAC_Pc2.Hydrophobicity.1), reveals that both high and low hydrophobicity values significantly influence cell penetration. This highlights the crucial role of hydrophobic distribution in mediating interactions between peptide and cell. Additionally, the Conjoint k-spaced Triad method, evaluating the occurrence of arginine and lysine patterns within the peptide sequence (KSCTriad_g5.g5.g5.gap1), significantly contributes to CPP prediction. Patterns showing frequent occurrences of arginine and lysine triplets separated by a single residue are strongly associated with CPP activity, emphasizing the importance of specific spatial arrangements of these charged amino acids in promoting penetration. The specific alanine amino acid composition (APAAC) is also informative for the classification between CPPs and non-CPPs. High APAAC values are predictive of CPP activity, suggesting that a higher and more even distribution of alanine can enhance membrane interaction and uptake. In contrast, lower APAAC values, indicative of sparse distribution and a less favorable hydrophobic-hydrophilic balance, correlate with non-CPP predictions. Similarly, the presence of valine (PAAC) critically influences CPP classification. Low valine levels lead to the positive classification of a penetrating peptide, suggesting minimal or localized presence of the previously mentioned amino acid. High concentrations, especially if unevenly distributed, are likely to hinder membrane penetration, underscoring valine's complex role in CPP functionality.

Three additional features emerged as important components in predicting CPP behavior: CKSAAP_IR.gap2, CKSAAP_LL.gap4, and CKSAAP_TR.gap4. In all cases, the previously mentioned features are related with the presence of amino acid pairs separated by specific gaps. The CKSAAP_IR.gap2 feature, representing the isoleucine-arginine pair separated by two residues, indicates that enhanced occurrences correlate with CPP propensity, whereas reduced occurrences align with non-CPP classification. This observation underscores the specific importance of isoleucine-arginine pairs with a gap of 2 residues for the penetrating capability. For the CKSAAP of the amino acid pair leucine-leucine separated by 4 residues, we observe a similar trend, where high values correspond to CPP prediction, while low values indicate non-CPP prediction. High values of CKSAAP_TR.gap4 feature, concerning the threonine-arginine pair separated by four residues, is associated with non-CPP predictions, suggesting that such spatial arrangements may inhibit cellular entry. Minimal occurrences of this pair yield no significant impact on the predictions. The previously discussed physicochemical patterns, with amino acids gaps of different length, underscore the intricate interplay between charge distribution, hydrophobicity, and spatial arrangement within peptide sequences that influence the peptide penetration capability across cellular membranes. The feature importance and impact of the training set are shown in Figure S1 - SI file. The feature importance analysis exhibits remarkable consistency between the training and testing sets, although with slight variations in their order. This consistency underscores the robustness of the identified features and their relevance across diverse peptide sequences.

## 3.4 Design Algorithm Analysis

The effectiveness of the design algorithm was evaluated using all the non-CPP sequences from the MLCPP2.0 testing set as input for the optimization loop. For this evaluation, the population and generation parameters were set to 500 and 50, respectively. In addition, the similarity distance was performed using the BLOSUM62 substitution matrix[64]. It is worth nothing that the users can modify the weights ($w_1$, $w_2$, and $w_3$) in the similarity distance, penetrability score, and local outlier factor terms of the fitness function depending on their specific optimization need. Increasing a weight above 1 will give greater impact to that term in the fitness function, and vice versa. The primary evaluation criteria include the amino acid changed pairs and types between the original and the optimized sequences.

The design of new penetrating sequences is achieved through the substitution of some amino acids in favor of others, aiming to maximize penetrability without deviating too much from the original peptide. The amino acids substitution pattern is shown through the heatmap matrix in Figure 3. The matrix entries were calculated by comparing the pairs of amino acids between the original and the optimized sequences. In addition, it is also shown the percentages of each deleted type of amino acids in Figure 3-right, as well as the percentages of each inserted type of amino acids in Figure 3-top. These two figures were obtained from the sum of each matrix's columns and rows, respectively. It is worth noting that the cationic amino acids (R and K) were the most inserted in the newly designed CPP. In detail, arginine is more substituted rather the lysine, which remarks its importance in the penetration capability of peptides. To emphasize this point, the arginine amino acid was almost never removed from the original sequences (Figure 3-right). Contrariwise, the anionic amino acids (D and E) were almost never newly inserted in the *de novo* designed CPP. This behavior is in agreement with the pivotal role exerted by the positive charge in the CPP. To reinforce this point, the aspartic and glutamic acids were often deleted from the original sequences, especially for the glutamic acid (Figure 3-right). Taken all together, such behavior can be related to the promotion of more positively charged sequence, since it is the most important feature of our ML model (Figure 2).

While positively charged amino acids are essential for the cell penetration ability of peptides, the hydrophobic amino acids and their distribution along the sequence also play an important role. Indeed, the hydrophobic amino acids alanine, cysteine, isoleucine, and leucine (A, C, I, and L) were highly inserted into the CPP optimized sequences (Figure 3-top). Interestingly, cysteine was moderately used as substitutive amino acid in the designed sequences and was also almost never deleted from the original sequences. This behavior suggests that cysteine can play a key role in conferring penetration capability to CPPs. Valine emerged as the most frequently substituted amino acid in the newly generated sequences, not only between the hydrophobic pool of amino acids, but between all of them. The feature importance and impact analyses (Figure 2) suggest that high content of valine in the sequence is associated with non-CPP prediction (PAAC_Xc1.V) and decreasing the valine content may shift the prediction more towards the CPP class. Figure 3 also highlights that the valine is predominantly replaced with isoleucine. This pattern requires further examination and will be explored more thoroughly in the discussion section. Another interesting observation arising from the analysis is the conspicuous addition and depletion of alanine, isoleucine, and leucine in the designed CPPs (Figure 3). This behavior indicates not only the importance of these specific amino acids, but also suggests that their distribution pattern within the sequence may play a crucial role for the penetration capability of CPPs. These findings underscore the specificity and diversity inherent in CPP design, leading to several optimization schemes that pursue the increase of penetration capability of the final peptide.

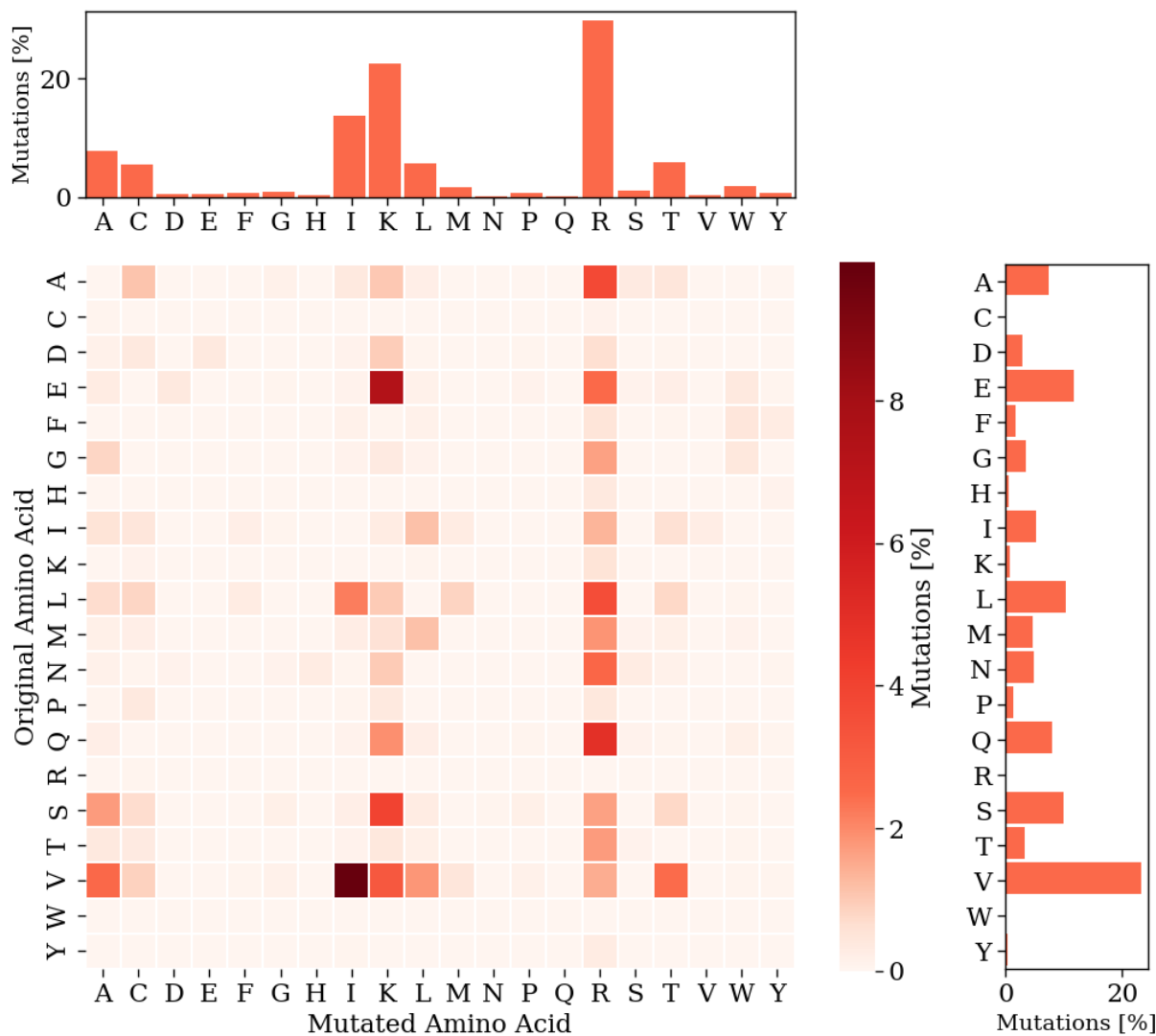

*Figure 3: Heatmap with the original replaced amino acids in the matrix's rows and the newly inserted amino acid in the matrix's columns. The results of the substituted amino acid pairs are shown as percentages. In addition, in the top figure is shown the percentages of each newly inserted type of amino acid, while in the right figure is shown the percentages of each deleted type of amino acid.*

We also performed the SHAP analysis on a single randomly selected peptide (IFSNTALVNCMRQTLQDTGHNP) before and after the optimization process (IFKRTALINCRRRTLQDTGHNP) (Figure 4). Other examples of SHAP analysis performed on randomly selected peptides are shown in Section S3 – SI file. It is interesting to note that the substituted amino acids after the optimization process shifted several features from non-CPP to CPP prediction, such as CTDC_charge.G1, CKSAAP_IR.gap2, APAAC_Pc2.Hydrophobicity.1, and PAAC_Xc1.V, among others.

The optimization algorithm increased the overall positive charged amino acids of the peptide through the addition of 3 arginine and 1 lysine residues, in substitution to a serine, asparagine, methionine and glutamine residues. These substitutions led to an increase of the CTDC_charge.G1 feature, which is crucial for cell penetration. This increase in charge is aligned with the previous observation of positively charged amino acids being essential for the CPP function. Furthermore, the algorithm strategically decreased the valine content (PAAC_Xc1.V), consistent with our earlier findings that a high valine content is associated with non-CPP prediction. Valine was substituted with isoleucine, a pattern that was highlighted in Figure 3 and supports the transition towards a CPP classification. In addition, the substitution of valine with isoleucine also contributed to an increase in the CKSAAP_IR.gap2 value, indicating a more favorable sequence-order correlation of these two amino acids for penetration ability. From another point of view, the design algorithm also focused on balancing the hydrophobicity-hydrophilicity property, as indicated by changes in the APAAC_Pc2.Hydrophobicity.1 feature.

The adjustments reflected in the SHAP value analysis, underscore the algorithm's ability to fine-tune specific sequence features to enhance CPP functionality. The importance of these modifications highlights the intricate relationship between amino acid composition, sequence order, and the biophysical properties required for cell penetration.

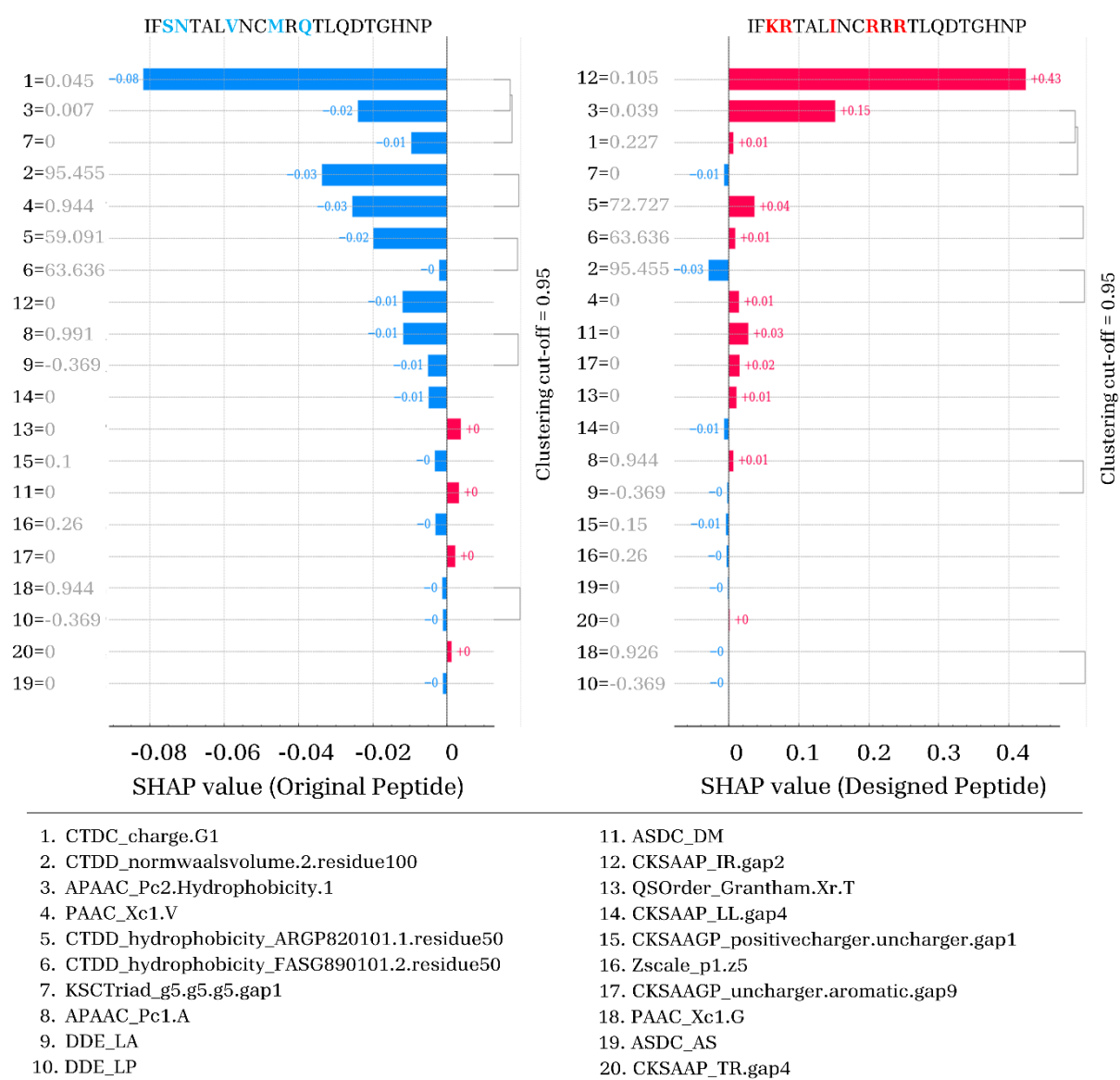

1. CTDC_charge.G1
2. CTDD_normwaalsvolume.2.residue100
3. APAAC_Pc2.Hydrophobicity.1
4. PAAC_Xc1.V
5. CTDD_hydrophobicity_ARGP820101.1.residue50
6. CTDD_hydrophobicity_FASG890101.2.residue50
7. KSCTriad_g5.g5.g5.gap1
8. APAAC_Pc1.A
9. DDE_LA
10. DDE_LP
11. ASDC_DM
12. CKSAAP_IR.gap2
13. QSOrder_Grantham.Xr.T
14. CKSAAP_LL.gap4
15. CKSAAGP_positivecharger.uncharger.gap1
16. Zscale_p1.z5
17. CKSAAGP_uncharger.aromatic.gap9
18. PAAC_Xc1.G
19. ASDC_AS
20. CKSAAP_TR.gap4

*Figure 4: SHAP values of a randomly selected peptide (IFSNTALVNCMRQTLQDTGHNP) before (on the left) and after (on the right) the optimization. In addition, it is highlighted in bolded blue the original deleted amino acid (top-left), and in bolded red the newly inserted amino acids (top right). In grey it is shown the value for each feature, before and after the optimization process.*

# 4 Discussion

Machine learning (ML) methods have emerged as powerful tools for the prediction of a wide range of biological phenomena. With the diffusion of extensive cell-penetrating peptides (CPPs) datasets, these techniques have found utility in predicting and classifying peptides with the ability to translocate across cellular membranes[35,36,38–40,43,46–48]. Despite the development of numerous ML models for predicting cellular uptake and membrane translocation of CPPs, several challenges remain in the field. To advance the design of novel CPPs, it is crucial to develop interpretable models that provide insights into the underlying factors driving CPP performance. Moreover, although ML predictions can be used for discriminating between CPPs and non-CPPs, they cannot be applied for the design of novel CPPs starting from a non-penetrating sequence. In this work, we have developed a pipeline named LightCPPgen, which utilizes a genetic algorithm (GA) empowered by a lightGBM-based predictor to generate and optimize non-penetrating peptides, enabling targeted modifications with minimal amino acid changes (Figure 1). In this framework, the main aim pursued by our work is to overcome the inherent challenges associated with traditional trial-and-error approaches of the experimental design. Similar methodologies have been explored in the literature, particularly in the context of designing novel antimicrobial peptides (AMPs)[65,66]. However, to the authors' knowledge, this is the first time an ML model on top of a GA was utilized to design novel CPPs starting from non-CPP sequences. This innovative approach specifically targets the candidate sequences' penetrability score, while trying to retain the original residues of the peptides to the greatest extent possible through the similarity distance. By maximizing similarity with the original non-penetrating macromolecule, the modified peptides are likely to retain the original biological and physicochemical properties that are crucial for their intended function, while also gaining the ability to cross cellular membranes. This minimally invasive modification approach is particularly valuable as it avoids the potential unintended effects that might arise from more radical alterations, thereby reducing the risk of detrimental impacts on peptide stability and functionality. Moreover, this strategy facilitates the easier integration of these peptides into existing therapeutic frameworks, enhancing their translational potential in clinical settings.

The lightGBM-based CPP predictor embedded in the optimization algorithm demonstrated robust performance in distinguishing between CPPs and non-CPPs, achieving high accuracy, specificity, and MCC (Table 1), while maintaining a high degree of efficiency. It is important to mention that the lightGBM-model performance metrics are in line with the one exhibited by MLCPP2.0[47] and SiameseCPP[48], while maintaining a high degree of efficiency through the use of only 20 features. Choosing the right model among those proposed, MLCPP 2.0, and SiameseCPP depends on the specific needs and consequences of classification errors in the application. Within the context of this work, where the main goal is to suggest new candidate CPPs to be synthesized in the laboratory while minimizing false positives and thus reducing resources and costs for synthesizing new peptides, the proposed classifiers might be preferable due to their higher specificity, overall accuracy, and MCC. This approach not only conserves laboratory resources but also accelerates the discovery and development of new CPPs by prioritizing the synthesis of the most promising candidates.

The feature importance analysis developed in this study has demonstrated the presence of key descriptors governing the predictive power of the ML model. The pivotal feature responsible for the penetrability score is the presence of a net positive charge on CPPs (CTDC_charge.G1, as shown in Figure 2). This result is in line with previous studies showing that this molecular property improves cellular translocation efficiency, enabling efficient delivery of therapeutic cargoes across biological barriers[67–70]. Furthermore, the positive charge of CPPs contributes to their ability to translocate cell membranes via various pathways, including direct penetration, macropinocytosis, and clathrin-mediated endocytosis, among the others[71–75]. It is worth mentioning that not only the absolute positive charge is an important parameter, but also its distribution (KSCTriad_g5.g5.g5.gap1 feature in Figure 2) in the peptide sequence[69,76]. The presence of asparagine, valine, glutamic acid, glutamine, isoleucine, and leucine amino acids at the C-terminal tale of the sequence, corresponding to high values of the CTDD_normwaalsvolume.2.residue100 feature, has been also identified as important factors leading the classification of peptides as non CPPs (Figure 2). The finding that a hydrophobic distribution pattern (APAAC_Pc2.Hydrophobicity.1, APAAC_Pc1.A and CKSAAP_LL.gap4 features in Figure 2) plays a crucial role in the penetration process aligns with recent research showing how incorporating a hydrophobic moiety into the arginine-rich CPPs is fundamental for increasing cell penetration[77]. We also found that CPP classification is also characterized by hydrophobic-cationic residues pair separated by 2 residues (CKSAAP_IR.gap2 in Figure 2). In summary, the interplay between hydrophobic and cationic moieties results to be a crucial characteristic for enhancing CPPs' ability to cross plasma membranes, in agreement with recent literature[77,78].

Leveraging a ML classifier, we designed a generative model based on a GA to systematically optimize CPP sequences. The effectiveness of the design algorithm in optimizing CPP sequences was evaluated by analyzing the amino acid changed pairs with a focus on their types and frequencies (Figure 3). As expected, the results indicate the insertion of positively charged amino acids (arginine-R and lysine-K) as leading factor driving the penetration propensity of the peptides. On the other hand, anionic amino acids like aspartic and glutamic acids (D and E), were both rarely inserted into the optimized peptide and frequently deleted from the original sequences. These observations align with the crucial role of positive charge in CPP functionality[67–70], as reflected in our ML model (section 3.3). In addition, our analysis highlighted the importance of hydrophobic amino acids, such as alanine, cysteine, isoleucine, and leucine (A, C, I, and L, respectively) in the CPP designing process (Figure 3). These hydrophobic residues were likely inserted into the CPP sequences, indicating their pivotal role in enhancing membrane penetration. Interestingly, the latter were also frequently deleted from the original sequences. This behavior suggests that their distribution pattern within the sequence may play a crucial role for the penetration capability of CPPs. It is worth nothing that the presence of hydrophobic amino acids in cationic-rich peptides is a key determinant in enhancing the penetrability capacity of CPPs[77,79–81]. Cysteine emerged as a moderately substituted amino acid, suggesting its potential in conferring penetration capability to CPPs (Figure 3), in line with recent studies showing cysteine rich penetrating peptides[82–84]. The valine (V)-isoleucine (I) substitution raised intriguing questions about the relationship between this specific amino acid substitution and the non-CPP/CPP peptides prediction (Figure 3). The feature importance and impact analyses (Figure 2) indicated that a high content of valine (PAAC_Xc1.V) is associated with non-CPP predictions, suggesting that reducing valine content may lead to classifications favoring the CPP class. It should also be noted that amino acid substitutions are influenced by the similarity distance evaluation, utilizing the BLOSUM62 substitution matrix. In this framework, the substitution of valine (V) with isoleucine (I) emerges as the most favorable, following the homologous V-V substitution. Given these observations, the prevalent V-I substitution may represent a potential limitation of the current algorithm. This aspect warrants further investigation in future studies to fully assess its biological meaning and impact.

## 5 Conclusions

This study introduces a novel ML-driven approach for de novo CPP design, leveraging the synergistic integration of predictive modelling and evolutionary algorithms. This approach offers a systematic and efficient strategy for generating and optimizing CPP sequences with robust performance and interpretability.

By elucidating key features governing CPP translocation capacity and optimizing algorithm parameters, we contribute to the advancement of rational CPP design and accelerate the development of peptide-based therapeutics with enhanced cellular penetration capabilities. Furthermore, this knowledge not only enhances our understanding of CPP biology but also guides the rational design of peptide-based therapeutics for various biomedical applications. In this context, our tool is designed with the overarching goal of reducing cost and time in wet laboratories. Moving forward, we envision further refinements and applications of our methodology to address evolving challenges in drug delivery and biomedical research.

## Acknowledgements

This work was supported by a grant from the Swiss National Supercomputing Centre (CSCS).

## Conflict of interest

The authors declare no competing interests.